\documentclass[aps,prl,floats,twocolumn]{revtex4}
\usepackage{graphicx}
\begin{document}


\newcommand{\co} {$^{59}$Co}
\newcommand{\na} {$^{23}$Na}
\newcommand{\naun} {Na$_{1-\epsilon}$CoO$_2$}
\newcommand{\nahuit} {Na$_{0.80}$CoO$_2$}
\newcommand{\etal} {{\it et al.}}
\newcommand{\ie} {{\it i.e.}}
\newcommand{\jpsj} {{J. Phys. Soc. Jpn.}}


\title{Non-magnetic insulator state in Na$_{1}$CoO$_2$ and phase separation of Na vacancies}

\author{C. de Vaulx$^1$, M.-H. Julien$^{1,*}$, C. Berthier$^{1,2}$, M. Horvati\'c$^{2}$, P. Bordet$^{3}$,
V. Simonet$^{4}$, D.P. Chen$^{5}$ and C.T. Lin$^{5}$}

\affiliation{$^1$Laboratoire de Spectrom\'etrie Physique,
Universit\'e J. Fourier \& UMR5588 CNRS, BP 87, 38402, Saint
Martin d'H\`{e}res, France}

\affiliation{$^2$Grenoble High Magnetic Field Laboratory, CNRS, BP
166, 38042 Grenoble Cedex 9, France}

\affiliation{$^3$Laboratoire de Cristallographie, CNRS, 38042
Grenoble Cedex 9, France}

\affiliation{$^4$Laboratoire Louis N\'eel, CNRS, BP 166, 38042
Grenoble Cedex 9, France}

\affiliation{$^5$Max-Planck-Institut for Solid State Research,
Heisenbergstrasse 1, 70569 Stuttgart, Germany}


\begin{abstract}

Crystallographic, magnetic and NMR properties of a Na$_x$CoO$_2$
single crystal with $x\simeq1$ are presented. We identify the
stoichiometric Na$_1$CoO$_2$ phase, which is shown to be a
non-magnetic insulator, as expected for homogeneous planes of
Co$^{3+}$ ions with $S=0$. In addition, we present evidence that,
because of slight average Na deficiency, chemical and electronic
phase separation leads to a segregation of Na vacancies into the
well-defined, magnetic, Na$_{0.8}$CoO$_2$ phase. The importance of
phase separation is discussed in the context of magnetic order for
$x$$\simeq$0.8 and the occurrence of a metal-insulator transition
for $x$$\rightarrow$1.

\end{abstract}
\maketitle
The discovery of superconductivity in H$_2$O-intercalated
Na$_{0.33}$Co$_2$~\cite{Takada03} has reopened investigations of
the physical properties of Na$_x$CoO$_2$ compounds, which remained
largely unexplored so far. In the simplest picture, the Co planes
contain $x$ non-magnetic ($S$=0) Co$^{3+}$ ions and 1-$x$
$S$=$\frac{1}{2}$ spins (Co$^{4+}$ low-spin state). One of the
surprises in the emerging phase diagram~\cite{Foo04} is thus the
presence of magnetic order at rather low Co$^{4+}$ concentrations
$x>0.75$. The origin of this order and how it evolves with $x$ are
presently unknown or
controversial~\cite{magneticorder,Sakurai04,Mendels05,Bayrakci05,Helme05,Carretta04}.
These questions are directly related to the electronic state of Co
and to the microscopic organization within CoO$_2$ layers. Their
elucidation is thus necessary for understanding properties such as
the remarkable thermopower of these oxides. As a matter of fact,
several hypothesis have already been raised in order to
rationalize magnetic order: peculiar Co$^{3+/4+}$ charge
order~\cite{Bayrakci05}, breakdown of the Co$^{3+/4+}$ ionic
picture~\cite{Helme05}, phase separation
\cite{Sakurai04,Carretta04} and/or intermediate ($S$=1) spin-state
for Co$^{3+}$~\cite{Bernhard04}.

In this context, a crucial question is whether the magnetic
transition temperature extrapolates to zero as $x$$\rightarrow$1,
and whether this occurs in a continuous way or not. Indeed,
Na$_1$CoO$_2$ is usually assumed to be a non-magnetic band
insulator with fully occupied $t_{2g}$ levels of Co$^{3+}$ ions.
This is supported by bulk susceptibility~\cite{Kikkawa86} and
transport measurements~\cite{Delmas81}. However, chemical phase
separation sometimes reported in this doping range may complicate
interpretations~\cite{Delmas81,Huang04}. Thus, investigation with
a local and bulk probe, such as nuclear magnetic resonance (NMR),
is clearly needed for characterizing electronic properties as
$x$$\rightarrow$1.

Here, we report NMR, X-Ray and magnetization measurements in a
single crystal of Na$_x$CoO$_2$ with $x\simeq 1$. The
stoichiometric Na$_1$CoO$_2$ phase is unambiguously identified and
is shown to be a non-magnetic insulator, as predicted. In
addition, due to slight Na deficiency in the single crystal, we
observe a distinct minority phase which has all the
crystallographic and magnetic properties of the $x=0.8$ phase,
including a magnetic transition at low temperature ($T$). We
connect these results with similar observations in
Li$_{x}$CoO$_2$~\cite{Menetrier99}, where phase separation might
be related to the first order character of the metal-insulator
transition at $x=0.95$ \cite{Marianetti04}. This new piece of
information on electronic properties of Na$_x$CoO$_2$ underlines
the need for microscopic theories taking into account magnetic and
metal-insulator transitions, as well as the possibility of phase
separation.

We used a piece ($3.5\times4.7\times0.45$ mm$^3$) of a larger
NaCoO$_2$ single crystal, grown in an optical floating zone
furnace~\cite{Chen04}. NMR spectra were obtained from the sum of
spin-echo Fourier transforms recorded at evenly spaced magnetic
field values. Quadrupolar interaction splits each \na~(resp. \co)
site into three (resp. seven) lines since the nuclear spin
$I$$=$$\frac{3}{2}$ (resp. $\frac{7}{2}$). Both $H||c$-axis and
$H||ab$-planes field orientations were investigated in order to
determine quadrupolar parameters. \na~shifts are given with
respect to the \na~resonance in NaCl. The reference for \co~shift
is 10.03~MHz~T$^{-1}$. Magnetic hyperfine shift $^{23,59}K$ data
for $H||c$ were corrected from the demagnetizing field effect:
$\Delta K_{c}^{\rm demag}$$\simeq$ 50 ppm (resp. 130~ppm) at
$T=300$~K (resp. 16~K). $T_1$ values were obtained after a comb of
$\frac{\pi}{2}$ pulses. \na~and \co~data were fitted with
appropriate formulas for magnetic relaxation and a stretched
exponent $\alpha$$\simeq$ 0.8-0.9. For the sharp \na~and
\co~signals, $\alpha$$\neq$ 1 is due to significant quadrupolar
contribution to the relaxation. For the broad \na~signal, the
relaxation is magnetic but $\alpha$$\neq$ 1 because of lines with
different $^{23}T_1$ values overlap.

\begin{figure}[t!]
\centerline{\includegraphics[width=3.7in]{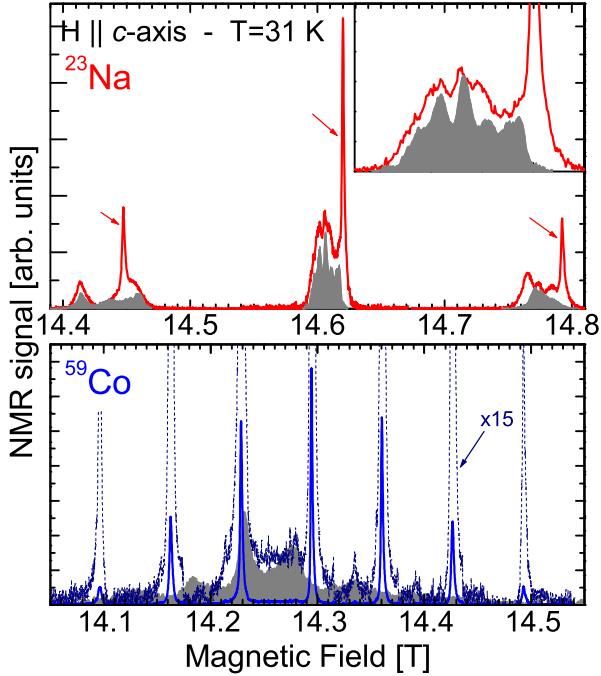}}
\vspace{-0.5cm} \caption{(Top) \na~NMR spectrum, showing the sharp
(arrows) and broad signals. Experimental conditions were chosen in
order to favor the broad signal. The sharp lines are thus
comparatively much more intense than shown here. The grey area is
the \na~NMR spectrum in a Na$_{0.8}$CoO$_2$ single
crystal~\cite{devaulx05}. Inset: detail of the \na~central lines
for $x=$1 and $x=$0.8 crystals. Note that the number of \na~NMR
lines and their shifts may somewhat vary from sample to sample
with $x\simeq0.8$ \protect\cite{devaulx05}, so that strict
one-to-one correspondence of \na~NMR lines is not required to
identify the $x\simeq0.8$ phase here. (Bottom) \co~NMR spectrum in
full scale (line) and in enlarged scale (dashed) in order to
evidence the broad \co~signal. Grey area: \co~NMR spectrum for
$x=$0.8~\cite{devaulx05}.}
\end{figure}

Fig.~1a shows a typical \na~NMR spectrum. The central line and
both satellite transitions can be clearly decomposed into two
contributions:

i) A sharp \na~signal, with quadrupolar coupling
$\nu_Q$=1.94(1)~MHz and the asymmetry parameter of the electric
field gradient tensor $\eta=|V_{xx}-V_{yy}|/V_{zz}=0$, in
reasonable agreement with room $T$ data~\cite{Siegel03}. The full
width at half-maximum (FWHM) at $T$=50~K is 17~kHz at 14.6~T and
13.4~kHz at 8.5~T, mostly due to nuclear dipolar broadening. Such
a very small magnetic and quadrupolar broadening indicates that
this single Na site must be in a very well ordered structure. The
magnetic hyperfine shift of this line is negligibly small:
$^{23}K_c^{\rm sharp}<35$~ppm over the whole temperature ($T$)
range (Fig.~2). This sharp \na~signal exactly corresponds to what
is expected for Na$_1$CoO$_2$: At variance with other
concentrations, this stoichiometric phase is characterized by a
single Na crystallographic site~\cite{structureNa1}, with axial
symmetry (thus $\eta$=0), and supposedly 100~\% Co$^{3+}$
occupation (thus a vanishing hyperfine field on \na).

ii) A broader signal, corresponding to a more disordered
structure, consists of at least three contributions. Quadrupolar
parameters ($\nu_Q=$1.78(5), 1.85(5) and 2.08(5)~MHz and
$\eta\simeq0$) are similar to the sharp signal. Since the relative
positions of the central lines scale with the magnetic field, the
second-order quadrupolar shift is negligible and the shift
differences between these lines are only due to the number of
different magnetic environments for \na~nuclei. All shift values
$^{23}K_c^{\rm broad}$ at any $T$ are remarkably close to those in
a Na$_{0.8}$CoO$_2$ single crystal~\cite{devaulx05} (Fig.~1), with
the the Curie-Weiss $T$ dependence typical of the the spin
susceptibility for $x>0.6$ samples (Fig.~2).

\begin{figure}[t!]
\centerline{\includegraphics[width=3.2in]{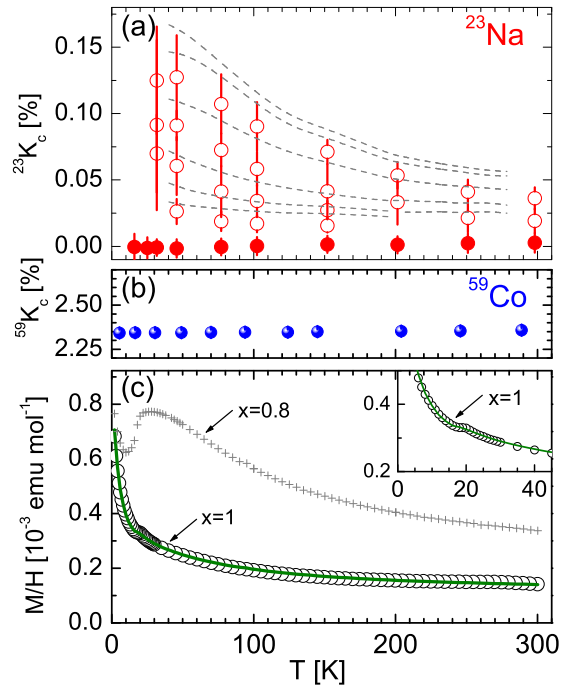}}
\vspace{-0.5cm} \caption{(a) \na~magnetic hyperfine shift for
$H||c$ ($^{23}$K$_c$) for the sharp (filled symbols) and broad
(open symbols) signals. Vertical bars are not error bars but the
FWHM. Dashed lines: $^{23}K_c$ in Na$_{0.8}$CoO$_2$
\cite{devaulx05}. (b) $^{59}$K$_c$ for the sharp \co~signal. FWHM
is close to the symbol size. (c) Symbols: magnetization data
($H=$1~T, $||c$). Line: fit explained in the text. Inset: same
data in enlarged scale showing a hump at $T_{\rm M}$=20~K.}
\end{figure}

Information on the magnetic fluctuations for the two sets of
signals can be obtained through site-selective measurements of
nuclear spin-lattice and spin-spin relaxation rates, $T_1^{-1}$
and $T_2^{-1}$, respectively. The two \na~signals clearly display
very different behaviors:

$^{23}T_1$ for the sharp line is extremely long, more than 20~s at
any $T$ (Fig.~3). Such long $^{23}T_1$ values indicate that the
spectral weight of magnetic fluctuations almost vanishes at the
nuclear Larmor frequency. This result is again very consistent
with a non-magnetic insulator, as expected for low-spin Co$^{3+}$
forming the Na$_1$CoO$_2$ phase. Indeed, the presence of magnetic
moments within CoO$_2$ layers would speed-up relaxation
noticeably. Since delocalized electronic wave functions would
significantly overlap with Na orbitals, a non-magnetic metallic
phase would also cause much faster relaxation with typically
$^{23}T_1^{-1}\propto T$, in contrast with our findings. The very
slow spin-lattice relaxation observed here is thus phase-specific.
As a matter of fact, such large $^{23}T_1$ values are not observed
on any \na~line in Na$_{0.8}$CoO$_2$~\cite{devaulx05}, despite the
large number of Co$^{3+}$ ions.

For the broad signal, on the other hand, both the magnitude and
the $T$ dependence of $^{23}T_1^{-1}$ are very close to those
measured on \na~lines in a Na$_{0.8}$CoO$_2$ single
crystal~\cite{devaulx05} (Fig.~3). This results nicely fits with
shift data attributing this signal to a Na-deficient phase.

The characteristic time $^{23}T_2$ of the spin-echo decay for the
sharp signal is roughly constant in the whole $T$ range (Fig.~3).
For the broad signal, however, the spin-echo decay shows a complex
behavior (to be analyzed in a forthcoming publication) below
$\sim$20~K, which is the hallmark of the magnetic ordered state in
Na$_{0.8}$CoO$_2$~\cite{devaulx05}. The occurrence of magnetic
order at $T_{\rm M}=20$~K definitely ensures that we are dealing
with the $x\simeq 0.8$ phase.

The \na~signal intensity, integrated over magnetic field values
and corrected for $T_1$ and $T_2$ effects, is proportional to the
number of nuclei in each phase. Here, we measure a ratio of the
broad to sharp \na~signals of 1:6 for both the central transition
and the satellites. This is thus considered as the volume ratio of
the two phases. Remarkably, this ratio is confirmed by
measurements of the bulk magnetization using a superconducting
quantum interferometer device (SQUID): $M/H$ data for $H||c$ can
be fitted with a constant term of 7.6~10$^{-5}$ emu.mol$^{-1}$, a
Curie-Weiss term $\frac{0.0031}{T+4}$ due to 0.8\% impurities in
the crystal, plus $\frac{1}{7}$ of the magnetization measured in a
Na$_{0.8}$CoO$_2$ single crystal (Fig.~2c). This last contribution
accounts for the small hump at $T_{\rm M}\simeq20$~K discernible
in the raw data.

Since $T_{\rm M}$ is known to be zero for $x<0.75$, 20-22~K for
$x=0.75-0.82$~\cite{magneticorder,devaulx05} and 27~K for
$x=0.85$~\cite{Mendels05}, we assume the Na concentration of the
Na-poor phase: $x_2\simeq 0.8-0.9$. With $x_2=0.8$ (resp. 0.9) and
the 1:6 intensity ratio, we estimate the average Na concentration
$x_{\rm av}= 0.97$ (resp. 0.98) in this single crystal.

\begin{figure}[t!]
\centerline{\includegraphics[width=3.6in]{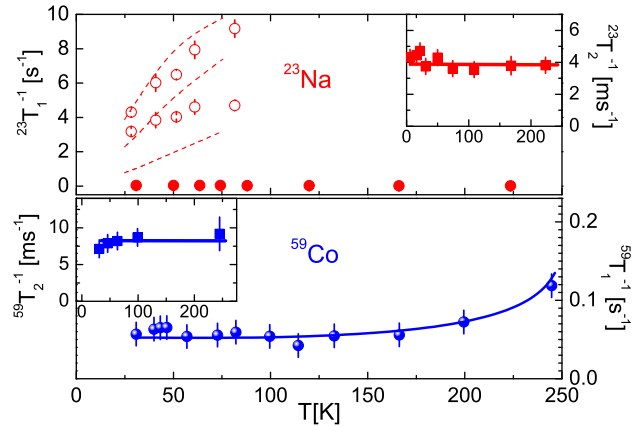}}
\vspace{-0.5cm} \caption{Top: $T_1^{-1}$ data for the broad (open
symbols) and sharp (filled symbols) \na~signals. For the later,
$T_1$ could be reliably extracted at two different positions only
on the broad spectrum, while $^{23}T_1$ in Na$_{0.8}$CoO$_2$
(dashes) could be measured for three resolved groups of
lines~\cite{devaulx05}. Inset: $T_2^{-1}$ for the sharp
\na~signal. Bottom: $T_1^{-1}$ and $T_2^{-1}$ (inset) for the
sharp \co~signal. Continuous lines are guides to the eye.}
\end{figure}

The \na~NMR results are also corroborated by \co~NMR: A typical
\co~NMR spectrum (Fig. 1b) is composed of the expected seven
lines, each of them being extremely sharp (FWHM of 15~kHz at
14.3~Tesla for the central line) and split by $\nu_Q$=0.68(2)~MHz.
This indicates a single Co site, again in a very well-ordered
crystallographic and electronic structure, with axial symmetry
derived from the asymmetry parameter $\eta$=0. The shift
$^{59}K^{\rm sharp}$ for this site is entirely attributed to the
$T$ independent orbital term $^{59}K_c^{\rm orb}$=2.35~\% and
$^{59}K_{ab}^{\rm orb}$=2.11~\%~(similar values are found for
Co$^{3+}$ sites in Na$_{0.7}$CoO$_2$, after correcting for the
reference choice)~\cite{Muka05}), {\it i.e.} the spin contribution
is negligible (Fig.~2). $^{59}T_1^{\rm sharp}$ is again extremely
long over the whole $T$ range (much longer than in
Na$_{0.3}$CoO$_2$~\cite{Ning04}) and $^{59}T_2^{\rm sharp}$ is
almost $T$-independent (Fig.~3). This sharp \co~signal, whose
properties are identical to those of the sharp \na~signal, is
undoubtedly attributed to the $x=1$ phase. The absence of
magnetism in CoO$_2$ layers is thus confirmed. Na$_1$CoO$_2$ is
the predicted band insulator. Note also that the long $^{23}T_1$
and $^{59}T_1$ values are indicative of the high purity of the
Na$_1$CoO$_2$ phase in our sample.

In addition to the sharp signal, the \co~NMR spectrum also
features a broad contribution (Fig.~1) with larger shift, which is
very reminiscent of the \co~signal in
Na$_{0.8}$CoO$_2$~\cite{devaulx05}. However, the weak integrated
intensity of this signal (consistent with the intensity ratio for
\na~signals) and its large broadening prevented us from accurate
relaxation measurements.

The sample was also characterized by X-ray diffraction using a
Seifert texture goniometer with Cu $\lambda K \alpha$ radiation.
The crystal was mounted in symmetric reflection geometry with the
scattering vector perpendicular to the crystal plate surface and
therefore expected to be parallel to the $c^*$-axis. A
$\theta$-$2\theta$ scan showed markedly split (00$\ell$)
reflections around $2\theta$=17$^\circ$ and 34$^\circ$, indicating
the existence of two phases in the crystal. Measurements of the
(003)$_{\rm R}$ and (006)$_{\rm R}$ reflections (rhombohedral unit
cell, or (002)$_{\rm H}$ and (004)$_{\rm H}$ for the corresponding
hexagonal cell) allowed to determine $c$-axis parameters. For
rhombohedral (resp. hexagonal) cells: $c_1$=16.044(5)~\AA~(resp.
10.696) for phase~1 is identical to the $c$ value in
Na$_{0.8}$CoO$_2$~\cite{Huang04}, and $c_2$=15.594(5)~\AA~(resp.
10.395) for the majority phase~2 is the value found in
Na$_1$CoO$_2$~\cite{structureNa1}. Thus, our single crystal shows
chemical phase separation between $x=0.8$ and $x=1$ phases, in
agreement with the results of Huang \etal~in
Na$_{0.89}$CoO$_2$~\cite{Huang04}.

It must be stressed that the observation of two phases is
different from trivial inhomogeneity, such as variation of the Na
content along the growth direction or gradual Na loss from the
crystal's surface. This would result in NMR line broadening, at
variance with the two distinct phases identified here. Our
observation of well-defined phases, together with the absence of
time evolution of NMR (and SQUID) results, demonstrate that NMR
probes the bulk of the single crystal and is therefore not
sensitive to possible surface degradation~\cite{sample}. This does
not exclude that the Na deficiency may partially result from
losses at the surface after the crystal synthesis, but our results
reveal that Na vacancies must segregate at some point.

Actually, several works suggest that Na$_x$CoO$_2$ compounds tend
to phase separate between stable, well-defined, Na
concentrations~\cite{Delmas81,Huang04,Huang05}. In particular, a
miscibility gap between $x=0.8$ and $x=1$ phases is suggested by
the results of Huang \etal~\cite{Huang04}. Such a chemical phase
separation would straightforwardly explain the electronic phase
separation evidenced here.

On the other hand, an important point to be considered is that
similar phase separation is clearly established in Li$_x$CoO$_2$,
which has the same CoO$_2$ layers as Na$_x$CoO$_2$ but different
Li stacking sequence. Experiments have demonstrated that
Li$_x$CoO$_2$ separates into Li-rich and Li-poor phases, over a
large concentration range $0.75\leq x \leq
0.94$~\cite{Menetrier99}. Furthermore, theoretical calculations
and the fact that the two phases have identical crystallographic
structure (unlike Na compounds) have led to the suggestion that
phase separation is not driven by Li vacancy
ordering~\cite{Ven98,Menetrier99}. Due to the concomitance of a
metal-insulator transition at $x=0.95$, electronic delocalization
is proposed as the driving force for phase
separation~\cite{Menetrier99}. According to Marianetti~\etal, this
transition is actually a Mott transition within the impurity band
of Li vacancies and its first order nature would explain
electronic phase separation~\cite{Marianetti04}. The high Li$^+$
mobility in this system would then allow global, chemical, phase
separation to proceed. Since there must be a metal-insulator
transition between $x=0.9$ and $x=1$ in Na$_x$CoO$_2$ as well,
similar physics is expected to be at work in this system. Another
remarkable fact is that phase separation is clearer in clean
Li$_x$CoO$_2$ samples than in impurity-doped ones
\cite{Tukamoto97}. This tends to reinforce the conclusion that
phase separation is an intrinsic phenomenon in these oxides.

Deciding whether the origin of phase separation is electronic or
chemical is beyond the scope of our work, which shows that both
occur. Identification of the sequence of stable phases is now
clearly needed and a possible dependence on the synthesis
conditions/methods should be clarified as well. Spatially
inhomogeneous electronic states as well as anomalous magnetism
might be the two key features to be considered for explaining
thermopower, which is actually strongest in this doping range.
These progresses should thus stimulate theoretical activity
towards microscopic models, specifically for the high doping
region from $x=0.75$ up to the metal-insulator transition.

Returning back to the magnetic properties of Na$_x$CoO$_2$, the
fact that the Na-poor phase has a concentration $x\simeq 0.8$ does
not support the hypothesis \cite{Sakurai04,Carretta04} that this
phase is itself phase separated. As to the proposal that a certain
amount of Co$^{3+}$ sites are in an intermediate spin state
($S=1$) \cite{Bernhard04}, no definitive conclusion can be drawn
from the present NMR results. Although our data rule out the
presence of $S=1$ spins in the $x=1$ phase, {\it local} changes of
the Co$^{3+}$ spin state close to local lattice deformations
(induced by a putative Co$^{3+}$/Co$^{4+}$ pattern) cannot be
excluded in $x\simeq0.8$ phases. However, it remains to be seen
whether magnetization data, which have been explained by a
concentration $x$ of $S$=$\frac{1}{2}$ spins~\cite{magneticorder},
could support $S$=$1$ moments as well.

We wish to thank H. Mayaffre for his constant support and fruitful
discussions.

Note added: Recently reported NMR spectra for the Na$_1$CoO$_2$
phase (G.~Lang \etal, cond-mat/0505668) quantitatively agree with
ours.


\begin{references}
\bibitem{sample} When not in the cryostat, the sample was stored in
liquid N$_2$ in order to prevent surface degradation in air.
\end{references}
\end{document}